\documentstyle[12pt,ams]{article}

\begin{document}

\thispagestyle{empty}
\begin{center}
\vspace{2cm}

{\Huge Calculating the Critical Temperature for Coleman-Weinberg GUTs
\\}

\vspace{1.5cm}

{\Large Richard Easther \\ William Moreau \\}

\vspace{1cm}

{\large
 Department of Physics and Astronomy, \\
 University of Canterbury, \\
 Private Bag, \\
 Christchurch, \\
 New Zealand. \\}

\vspace{1cm}

{\large email: \\
 rje@phys.canterbury.ac.nz \\
 w.moreau@canterbury.ac.nz \\}

\end{center}

\vspace{1.5cm}

\noindent PACS: 1210

\noindent Preprint: NZ-CAN-RE-92/1

\noindent To appear in J. Phys. G

\newpage

\pagenumbering{arabic}
\pagestyle{myheadings}

\section*{Abstract}

We study the finite temperature effective potential of the Higgs scalar in
GUTs with Coleman-Weinberg symmetry breaking. The critical temperature is
derived without employing a high temperature approximation to the effective
potential, and the limitations of such approximations are discussed.

\section{Introduction}

This paper takes another look at the determination of the critical temperature
for Coleman-Weinberg models \cite{Coleman ET1973a,Dolan ET1974a}. To do this
we must examine the one loop finite temperature effective potential,
$V(\phi,T)$, of the Higgs field that brings about symmetry breaking in the
GUT.

When $T \neq 0$, $V(\phi,T)$ typically has two minima: one at the origin
and one at some non-zero value of $\phi$, although at very high
temperatures this second minimum vanishes. The critical temperature,
$T_{c}$, is defined to be the temperature at which the two minima are
degenerate \cite{Brandenberger1985a}. At temperatures less than $T_{c}$,
spontaneous symmetry breaking becomes energetically favourable, but in
models of inflation based on Coleman-Weinberg GUTs \cite{Albrecht
ET1982a,Linde1982a,Olive1990a}, considerable supercooling will occur
before symmetry breaking takes place. To form a clear picture of the
evolution of the primordial universe in the context of such a theory it
is important that the value of $T_{c}$ is accurately known.

For a specific model, $T_{c}$ can be computed numerically without
recourse to a high temperature approximation to the effective potential.
However, a numerical calculation lacks versatility as it pertains to a
specific set of parameter values and obscures the general form of the
solution.

Approximate forms of the effective potential have been used to derive
algebraic expressions for $T_{c}$ in terms of the parameters of the GUT.
Unfortunately, these calculations must be taken with a grain of salt
since high temperature approximations to $V(\phi,T)$ are unreliable in
the vicinity of $T_{c}$. In fact, as Brandenberger
\cite{Brandenberger1985a} points out, this method leads to a result that
is no more informative than simple dimensional analysis.

In this paper, $T_{c}$ is calculated analytically to one loop level,
without using the high temperature approximation. In the process, we
obtain a concise derivation of the high temperature approximation to the
effective potential and discuss the circumstances in which such an
approximation is applicable.

\section{The finite temperature effective potential}

Consider a generic GUT with Coleman-Weinberg symmetry breaking.
Including gauge loops only, the one loop finite temperature effective
potential of the Higgs field, $\phi$, in such a theory is \cite{Billoire
ET1982a}
\begin{equation}
V(\phi ,T) = B\phi ^{4} \left[ \ln\left( \frac{\phi^{2}}{\sigma^2} \right)
- \frac{1}{2} \right] + \frac{DT^{4}}{2\pi ^2} I(\tau) ,
\label{Vdef}
\end{equation}
where
\begin{equation}
I(\tau) = \int_{0}^{\infty} dx\mbox{ }x^{2} \ln{\left[ 1 - \exp{\left(
-\sqrt{ x^{2} + \tau^{2}}\right)}\right]} ,    \label{Iint}
\end{equation}
and
\begin{equation}
\tau = \frac{M\phi}{T} , \makebox[10mm]{}  B = \frac{3n M^{4}}{64\pi
^{2}} , \makebox[10mm]{}  D = 3n .
\label{coeffs}
\end{equation}
At zero temperature, the minimum of $V(\phi,T)$ is located at $\phi =
\sigma$. The number of heavy gauge bosons is $n$ and M is a coefficient
determined by the gauge group and the coupling constant. In the case of
$SU(5) \rightarrow SU(3) \times SU(2) \times U(1)$
\cite{Billoire ET1982a},
\begin{equation}
M^2 = \frac{25}{8} g^{2} ,
\end{equation}
where $g^{2}/4\pi$ is the coupling constant.

The temperature dependent part of the effective potential, $I(\tau)$,
must be expressed in a more tractable form. Expanding the logarithm in
equation~(\ref{Iint}), integrating by parts and employing (3.479~1) of
Gradshteyn and Ryzhik \cite{Gradstheyn ET1980a} gives
\begin{equation}
I(\tau) = -\sum_{m = 1}^{\infty} \frac{\tau ^{2}}{m^{2}} K_{2}(m\tau) ,
\label{Ibess}
\end{equation}
where $K_{2}$ is a modified Bessel function of the second kind
\cite{Watson1966a}. Bollini and Giambiargi \cite{Bollini ET1984a} use
dimensional regularization to derive the finite temperature effective
potential in an arbitrary number of dimensions and Konoplich
\cite{Konoplich1989a} obtains the same result with zeta function
methods. The above series can be extracted immediately from their work
but the derivation given here has the advantage of being more direct.

The series in equation (\ref{Ibess}) converges rapidly but it is useful
to write it in a form where its dependence on $\tau$ is more
transparent. Using a little-known result of Watson \cite{Watson1931a},
$I(\tau)$ becomes
\begin{eqnarray}
I(\tau) &=&
-\frac{\pi}{3}\sum_{m = 1}^{\infty}\left[ \left(\tau ^ 2 + 4\pi
^{2}m^{2}\right)^{\frac{3}{2}} - (2m\pi )^{3} - 3m\pi \tau ^{2} -
\frac{3\tau ^{4}}{16m\pi} \right] \nonumber \\
& &\mbox{} - \frac{\pi ^{4}}{45} + \frac{\pi ^2}{12}\tau ^2 -
\frac{\pi}{6}\tau ^3 - \frac{\tau ^{4}}{16}\left(\gamma - \frac{3}{4} +
\ln{\frac{\tau}{4\pi}}\right) .
\label{Isum}
\end{eqnarray}
When $\tau < 2\pi$, the binomial theorem can be used to re-express the
infinite series in the above equation as a power series, giving
\begin{eqnarray}
I(\tau) &=&
-\frac{\tau^{4}}{8}\sum_{m = 1}^{\infty}
(-1)^{m}\frac{(2m-1)!!}{2^{m}(m+2)!}
\left(\frac{\tau}{2\pi}\right)^{2m}\zeta (2m+1)    \nonumber \\
& &\mbox{} - \frac{\pi ^{4}}{45} + \frac{\pi ^2}{12}\tau ^2 -
\frac{\pi}{6}\tau ^3 - \frac{\tau ^{4}}{16}\left(\gamma - \frac{3}{4} +
\ln{\frac{\tau}{4\pi}}\right) .
\label{Isum2}
\end{eqnarray}
The high temperature approximation to the effective potential is
obtained from equation (\ref{Isum2}) by taking the leading term(s) in
$\tau$. In particular, at very high temperatures only the term in
$\tau^{2}$ is retained. Inspection of the higher order terms shows that
for $\tau \gtrsim 1$, some of the discarded terms are equal in magnitude
to, or larger than, those that are kept. Therefore, approximations for
$I(\tau)$ are unreliable unless $\tau \ll 1$.
\section{Calculating the Critical Temperature}
In order to calculate the critical temperature, the value of $\phi_{c}$,
the location of the minimum of $V(\phi,T)$ when $T = T_{c}$ must be
known. By definition, at the critical temperature the minimum of
$V(\phi,T)$ is degenerate with the minimum at the origin. This gives two
equations that can be solved simultaneously for $\phi_{c}$ and $T_{c}$,
\begin{eqnarray}
V(0,T_{c}) &=& V(\phi_{c} ,T_{c}) \label{Vcond}\\
\left. \frac{d}{d\phi} V(\phi,T) \right|_{\phi = \phi_{c}} &=& 0 .
\label{dVcond}
\end{eqnarray}
Substituting the generic form of the effective potential,
equation~(\ref{Vdef}), into equation~(\ref{Vcond}) gives
\begin{eqnarray}
 \frac{B}{D}\frac{\phi ^{4}}{T^{4}} \left[ \ln\left(
\frac{\phi^{2}}{\sigma^2} \right) - \frac{1}{2} \right] + \frac{\tau ^{2}}{24}
-
\frac{\tau ^{3}}{12\pi} - \frac{\tau ^{4}}{32\pi ^2}\left(\gamma -
\frac{3}{4} + \ln{\frac{\tau}{4\pi}}\right)  & & \nonumber \\
\mbox{}  -\frac{\tau ^{4}}{16\pi^{2}}\sum_{m = 1}^{\infty}
(-1)^{m}\frac{(2m-1)!!}{2^{m}(m+2)!}
\left(\frac{\tau}{2\pi}\right)^{2m}\zeta (2m+1) &=& 0
\label{Vzero}
\end{eqnarray}
where we have anticipated that $\tau_{c} < 2\pi$ and used
equation~(\ref{Isum2}).
{}From equation~(\ref{dVcond}) we obtain (after multiplying by $\phi /4 $)
\begin{eqnarray}
\frac{B}{D} \frac{\phi ^{4}}{T^{4}} \ln{\left(\frac{\phi^{2}}{\sigma^2}
\right)} + \frac{\tau ^{2}}{48} - \frac{\tau ^{3}}{16\pi} - \frac{\tau
^{4}}{32\pi ^2}\left(\gamma - \frac{1}{2} +
\ln{\frac{\tau}{4\pi}}\right)  & & \nonumber \\
\mbox{} -\frac{\tau ^{4}}{32\pi^{2}}\sum_{m = 1}^{\infty}
(-1)^{m}\frac{(2m-1)!!}{2^{m}(m+1)!}
\left(\frac{\tau}{2\pi}\right)^{2m}\zeta (2m+1) &=& 0 .
\label{dVzero}
\end{eqnarray}
The logarithmic terms are eliminated from these equations by subtracting
equation~(\ref{dVzero}) from equation~(\ref{Vzero}) and then using
equation~(\ref{coeffs}). This gives
\begin{eqnarray}
\frac{\tau}{\pi} - 6\sum_{m = 1}^{\infty}
(-1)^{m}\frac{m(2m-1)!!}{2^{m}(m+2)!}
\left(\frac{\tau}{2\pi}\right)^{2m + 2}\zeta (2m+1) &=& 1 .
\end{eqnarray}
The solution of this equation is $\tau_{c}$, the value of $\tau$ when the two
minima are degenerate. Since it is a power series in $\tau$, it can inverted by
standard methods \cite[\S 4.5]{Morse ET1953a}, to get
\begin{equation}
\frac{\tau_{c}}{2\pi} = \frac{1}{2} + \sum_{n=4}^{\infty} b_{n} .
\label{taucrit}
\end{equation}
The terms $b_{n}$ are given by
\begin{equation}
b_{n} = \frac{1}{n2^{n}} \sum_{d,e,f,\cdots}(-1)^{d + e + f +
\cdots} \frac{n\cdots(n-1+d+e+f+\cdots)}{d!e!f!\cdots}
(a_{4})^{d}(a_{6})^{e}(a_{8})^{f} \cdots
\end{equation}
with the sum running over all $d,e,f,\cdots$ such that $3d + 5e + 7f +
\cdots = n-1$.
The $a_{n}$ are:
\begin{equation}
a_{2m+2} = (-1)^{m+1} \frac{3m (2m-1)!!}{2^{m}(m+2)!} \zeta (2m+1) .
\end{equation}
This is an exact result and will, in principle, yield $\tau_{c}$ to an
arbitrary degree of precision. The first few terms are
\begin{equation}
\frac{\tau_{c}}{2\pi} \approx \frac{1}{2} - \frac{\zeta(3)}{64}\left[ 1 -
\frac{\zeta(3)}{8} + \frac{15 \zeta(5)}{256} \right] +
\frac{3 \zeta(5)}{1024} - \frac{9 \zeta(7)}{16384}
\end{equation}
giving $\tau_{c} = 3.05$.

We are now in a position to calculate $\phi_{c}$. From equations~(\ref{Ibess})
and (\ref{dVcond}),
\begin{eqnarray}
\phi_{c}^{2} &=& \sigma^2 \exp{\left[
-\frac{8}{\tau_{c}}\sum_{n=1}^{\infty} \frac{1}{n} K_{1}(n\tau_{c})
\right]} \\
&=& \frac{\sigma^{2}}{1.106} . \nonumber
\end{eqnarray}
Finally, we obtain the critical temperature:
\begin{equation}
T_{c} = \frac{M\phi_{c}}{\tau_{c}} .
\label{Tc}
\end{equation}

Since $\tau_{c} > 1$, the high temperature approximation to the effective
potential is not reliable when $T \approx T_{c}$ and $\phi \approx \phi_{c}$.
Hence, calculations of $T_{c}$ based on such an approximation cannot be
trusted.

As an example, consider $SU(5)$ breaking to $SU(3) \times SU(2)
\times U(1)$ with $g^{2}/4\pi = 1/40$ and $\sigma = 4.5 \times 10^{14}
GeV$. In this case, $T_{c} = 1.39 \times 10^{14} GeV$.  For comparison, a high
temperature approximation \cite{Brandenberger1985a} gives $T_{c} \approx 6
\times 10^{13} GeV$, which differs by a factor of two from the value obtained
here.

However, if the running gauge coupling is used \cite{Sher1981a}, $M$
has a logarithmic dependence on the temperature so we cannot assign it a
constant value. For $SU(5) \rightarrow SU(3) \times SU(2)
\times U(1)$ the one loop running coupling is
\begin{equation}
\frac{g^{2}}{4\pi} = \frac{3\pi}{10 \log{\frac{T^{2}}{\Lambda ^{2}}}},
\label{grun}
\end{equation}
where $\Lambda = 2.74 \times 10^{6} GeV$. Let $\hat{T}_{c}$ be the
critical temperature calculated using a running coupling. From
equation~(\ref{Tc}) and the definition of $M^{2}$,
\begin{equation}
\hat{T}_{c}^{2} \log{\frac{\hat{T}_{c}^{2}}{\Lambda^{2}}} =
\frac{15\pi^{2}}{4} \frac{\phi_{c}^{2}}{\tau_{c}^{2}}.
\label{T-Hat}
\end{equation}
We obtain $\hat{T}_{c}$ in terms of $T_{c}$, the value of the critical
temperature calculated assuming a constant coupling. Define $\delta$ by
\begin{equation}
\hat{T}_{c}^{2} = (1+\delta)T_{c}^2  .
\label{deltadef}
\end{equation}
If the value of $g^{2}/4\pi$ used to obtain $T_{c}$ was reasonable, then
$\delta$ may be estimated by substituting equation~(\ref{deltadef}) into
equation~(\ref{T-Hat}). Expanding the logarithm in $\delta$, and retaining
only the first power gives
\begin{equation}
 \left[ 1 + \log{\frac{T_{c}^2}{\Lambda^{2}}} \right] \delta  =
\frac{15 \pi^{2}}{4} \left( \frac{\phi _{c}}{\tau_{c} T_{c}}
\right)^{2} - \log{\frac{T_{c}^2}{\Lambda^{2}}},
\end{equation}
which yields $\hat{T}_{c} = 1.43 \times 10^{14} GeV$.

\section{Discussion}

At the critical temperature, $T_{c}$, we find that $\phi_{c}$, the position of
the minimum of the effective potential when $T = T_{c}$, is related to the
zero temperature minimum, $\sigma$, by a multiplicative constant that is
independent of the model's parameters. The critical temperature is expressed
concisely in terms of $\tau_{c}$, a dimensionless constant which does not
depend on the particular parameter values of the Coleman-Weinberg model being
considered.

This paper calculates $T_{c}$ for a Coleman-Weinberg model to one loop order
without making use of a high temperature approximation to the effective
potential. The contribution of higher loops to the value of $T_{c}$ is
expected to be of the order $e^{2}T_{c}$ where $e^{2}$ is the coupling between
gauge bosons and the Higgs \cite{Sher1989a}. In order for perturbation theory
to be valid, this quantity must be much less than $T_{c}$, so the correction
to $T_{c}$ from higher loops will be small.

As Coleman and Weinberg \cite{Coleman ET1973a} note, higher loop corrections
can have the effect of converting the local maximum at the origin at $T = 0$
into a local minimum. However, since no graphs can affect the value of the
temperature independent portion of $V(\phi,T)$ at $\phi = 0$,
equation~(\ref{Vcond}) is unaffected by adding higher order terms.

Other definitions of the critical temperature exist in the literature. Dolan
and Jackiw \cite{Dolan ET1974a} define it as the temperature at which broken
symmetry first becomes possible, that is when a local minimum appears in
$V(\phi,T)$ at $\phi \neq 0$. In contrast, we have defined $T_{c}$ to be the
temperature at which $\phi = 0$ is no longer the global minimum of $V(\phi,T)$
and spontaneous symmetry breaking becomes energetically favourable. Using
Dolan and Jackiw's defintion of $T_{c}$, $V(\phi,T)$ has a point of inflexion
at the critical temperature and finding it requires us to solve for the values
of $\phi$ and $T$ for which the first and second derivatives of $V(\phi,T)$
are zero. These equations are similar to those analysed in this paper and can
be solved using the same techniques. We have carried out this calculation and
it yields the result that the effective potential has a point of inflexion for
$\tau = 2.055$. For $SU(5) \rightarrow SU(3) \times SU(2) \times U(1)$ this
value corresponds to a critical temperature of $1.67 \times 10^{14} GeV$.

Finally, when $T \approx T_{c}$ and $\phi \approx \phi_{c}$ we have shown that
the high temperature approximation is unreliable for any Coleman-Weinberg
model of the form given by equation~(\ref{Vdef}). Thus, $T_{c}$ cannot be
calculated accurately with an approximation to the temperature dependent
portion of the effective potential. However, we have obtained a concise
expression for $T_{c}$ in terms of $\tau_{c}$ and the parameters of the
Coleman-Weinberg model which is exact to one loop order and allows the
critical temperature to be obtained immediately for any specific model.


\newpage
\begin{thebibliography}{99}

\bibitem{Coleman ET1973a} Coleman S and Weinberg E (1973) Phys. Rev. D
{\bf 7}  1888-1910

\bibitem{Dolan ET1974a} Dolan L and Jackiw R (1974)  Phys. Rev. D
{\bf 9} 3320-41

\bibitem{Brandenberger1985a}  Brandenberger R (1985) Rev. Mod. Phys.
{\bf 57}  1-60

\bibitem{Albrecht ET1982a}  Albrecht A and Steinhardt P (1982) Phys. Rev.
Lett. {\bf 48} 1220-23

\bibitem{Linde1982a}  Linde A (1982) Phys. Lett. B {\bf 108}  389-93

\bibitem{Olive1990a}  Olive K (1990) Phys. Rep. {\bf 190}  307-403

\bibitem{Billoire ET1982a}  Billoire A and  Tamvakis K (1982) Nuc. Phys. B
{\bf 200} 329-44

\bibitem{Gradstheyn ET1980a}  Gradshteyn I and  Ryzhik I 1980 {\em Table of
Integrals, Series and Products (Corrected and Enlarged Edition)\/}
(San Diego: Academic Press)

\bibitem{Watson1966a}  Watson G (1966) {\em A Treatise on the Theory of Bessel
Functions, 2nd ed\/} (Cambridge: University Press)

\bibitem{Bollini ET1984a}  Bollini C and  Giambiargi J (1984) Phys.
Lett. B {\bf 134}  436-8

\bibitem{Konoplich1989a} Konoplich R (1989) Theor. Mat. Phys. {\bf 78}
315-25

\bibitem{Watson1931a} Watson G (1931) Quart. J. of Math. (Oxford) {\bf 2}
298-308

\bibitem{Morse ET1953a} Morse P and Feshbach H (1953)  {\em Methods of
Theoretical Physics, Vol. 1\/} (New York: McGraw Hill)

\bibitem{Sher1981a} Sher M (1981) Phys. Rev. D {\bf 24}  1699-1701

\bibitem{Sher1989a} Sher M (1989) Phys. Rep. {\bf 179}   273-418


\end{thebibliography}
\end{document}